\documentclass[pra,superscriptaddress,showpacs,nofootinbib,twocolumn]{revtex4}
\usepackage{ulem}
\usepackage{amsmath}
\usepackage{amstext}
\usepackage{latexsym}
\usepackage{graphicx}
\usepackage{amsfonts}
\usepackage{epsfig}

\usepackage{color}
\newcommand{\petr}[1]{\textcolor{black}{#1}} 

\newcommand{\Tr}{\mathrm{Tr}}

\begin{document}
\title{Vacuum as a less hostile environment to entanglement}

\author{Petr Marek}

\affiliation{School of Mathematics and Physics, The Queen's University,
  Belfast BT7 1NN, United Kingdom}

\author{Jinhyoung Lee}

\affiliation{Department of Physics and Quantum Photonic Science Research
  Center, Hanyang University, Sungdong-Gu, 133-791 Seoul, Korea}

\author{M. S. Kim}

\affiliation{School of Mathematics and Physics, The Queen's University,
  Belfast BT7 1NN, United Kingdom}

\date{\today}

\begin{abstract}
We derive sufficient conditions for infinite-dimensional systems whose entanglement is not completely lost in a finite time during its decoherence by a passive interaction with local vacuum environments. The sufficient conditions allow us to clarify a class of bipartite entangled states which preserve their entanglement or, in other words, are tolerant against  decoherence in a vacuum. We also discuss such a class for entangled qubits.
\end{abstract}
\pacs{03.65.Yz, 03.67.Mn,42.50.Dv} \maketitle

\section{Introduction}
\label{sec:i}

Quantum coherence, which is due to a fixed relative phase between wave functions of a quantum system, may be manifested in a unipartite or a multipartite system.  Various types of nonclassical properties, which have their origin in quantum coherence, have been discussed for a unipartite system, particularly, in the context of quantum optics~\cite{Barnett}.  Quantum coherence in a multipartite system can give a strong correlation between particles, which cannot be explained by classical theory~\cite{EPR,Bell64}.  This, so-called entanglement, is a key ingredient for quantum
information protocols such as quantum cryptography~\cite{cryptography},
teleportation~\cite{teleportation}, and computing~\cite{computing}. When a quantum system is embedded in the real world environment, it is known that decoherence degrades the entanglement even more rapidly than the quantum coherence of a unipartite system.   It
is thus important to study the decoherence of entanglement in order to find ways of circumventing it.

It has been shown that entanglement bears some connection to nonclassicality manifested by a unipartite system.  Indeed,  in order to see entanglement in the Gaussian output fields of a beam splitter, at least one of the two input field modes\footnote{Throughout the paper, the word `{\it mode}' is used to specify a light field instead of `$\cdots${\it partite system}'.} has to be squeezed (squeezing is a nonclassical nature of a field)~\cite{nonclassicality}.  When an antibunched field (antibunching is another nonclassical nature of a field) and the vacuum are input to a beam splitter, the output fields have also been shown to be entangled~\cite{nonclassicality1}. 

Quantum coherence manifested in a unipartite system or multipartite correlations suffers from
the effects of decoherence. Adding thermal noise with the amount of two
units of vacuum fluctuations rules out any nonclassicality, initially
imposed in the single mode field~\cite{nonclassicality2}. However, when a
nonclassical field is embedded in a vacuum environment via linear
and passive interaction (pure loss), it takes infinite time for the
nonclassicality to disappear completely~\cite{nonclassicality2}.  This is also true for some
entangled states which have been generated from Gaussian nonclassical states with linear devices~\cite{Duan}. In
a sense, one may say that nonclassicality and entanglement are rather
``tolerant against the decoherence in a vacuum environment.''

Recently, it was pointed out~\cite{Yu,Santos} that the tolerance in a vacuum is
not universal for entanglement; for some entangled qubits, pure or
impure, their entanglement vanishes at ``finite'' time even when they
are in a vacuum. This has also been demonstrated experimentally~\cite{Almeida}. It
is likely that a certain class of entangled states are tolerant in a
vacuum whereas others are disentangled earlier, which has been dubbed as ``sudden death
of entanglement'' (SDE).  

As the entanglement is a key resource of quantum information, the finite time disentanglement is clearly an unwanted effect. Since the passive interaction with a local vacuum bath, in the form of loss and inefficient detection, is an intrinsic part of quantum information protocols relying on quantum optics, it is useful to specify state parameters and interaction strengths when there is no danger of SDE's occurring.
In this paper we attempt to clarify a class of entangled states that are
tolerant against the decoherence in local vacuum environments. For this
purpose we derive sufficient conditions for infinite-dimensional systems
whose entanglement is not completely destroyed in finite time during the
decoherence in vacuum. The sufficient conditions enable us to clarify a
class of bipartite entangled states which preserve their entanglement, in other words, are tolerant to a vacuum decoherence.
We then briefly consider two-qubit systems and, with help of both our criteria and a qubit entanglement measure, discuss their behavior with respect to vacuum decoherence.

\section{Decoherence model}
\label{sec:ii}

Suppose that subsystem $a$ of a multipartite system $s$ interacts {\em
  linearly and passively} with a vacuum environment $r$. In this
physical situation, subsystem $a$ loses its energy to the
vacuum. Such vacuum decoherence is
described by a dynamic evolution of the density matrix $\rho_{s}$ for
the whole system $s$. If the system is initially in state $\rho_s(0)$ and its subsystem $a$ starts to interact with the vacuum in
$|0\rangle_r$, then the evolved state $\rho_{s}(t)$ at a
certain interaction time $t$ is given by
\begin{equation}\label{decoh1}
  \rho_{s}(0)\rightarrow \rho_{s}(t)= \Tr_r[U_{ar}
  \left( \rho_s(0)\otimes
    |0\rangle_r {}_r\langle 0| \right) U_{ar}^{\dag}],
\end{equation}  
where $U_{ar}$ stands for the unitary operation coupling subsystem
$a$ and the vacuum $r$. The relation in Eq.~(\ref{decoh1}) describes the
single-channel decoherence when subsystem $a$ interacts with its vacuum
environment. Decoherence of additional channels  can be straightforwardly
incorporated by employing series of the corresponding single-channel
decoherences, as long as the subsystems interact with an independent
environments. We also note that, even though the environment responsible
for the decoherence consists of many modes, a single-mode bath can
equivalently describe the decoherence if a time-dependent coupling
replaces the coupling constant~\cite{lee04}. The unitary operation $U_{a
  r}$ of the passive linear interaction transforms annihilation
operators of respective modes $a$ and $r$ as~\cite{carmichael}
\begin{equation}\label{decoh2}
  a \rightarrow \sqrt{\eta} a + \sqrt{1-\eta}e^{i\phi} r
\end{equation}
up to some overall phase factor. Here, $\eta$ represents the time-dependent coupling between the system and the environment.  When the system is fully assimilated to the environment, $\eta=0$.  When the system is in its initial state, $\eta=1$. For optical systems, this decoherence
implies a linear loss which is caused, for example, by unwanted
absorption and/or reflection of optical components, imperfect
mode-matching or inefficient detection.

Another elegant way to obtain the decohered quantum state is to employ the
Kraus representation. In this approach, the evolved density matrix is
given in the form\\
\begin{equation}\label{decoh3}
  \rho_s (t) = \sum_{n=0}^{\infty} K_n \rho_s (0) K_n^{\dag}.
\end{equation}   
The Kraus operators $K_n$ satisfy the completeness relation of $\sum_n
K_n^{\dag}K_n = \openone$ and they are obtained from Eq.~(\ref{decoh1})
as
\begin{equation}
  K_n = \langle n|_r U_{ar}|0\rangle_r =  \sqrt{\eta}^{a^{\dag} a}
  \frac{ (a\sqrt{1-\eta}e^{i\phi})^n}{\sqrt{n!}},
\end{equation}
where $a^\dag$ is the creation operator for mode $a$.

\section{General scenario}
\label{sec:iii}

The decoherence in vacuum, discussed in Sec.~\ref{sec:ii}, is fairly
benign for single-mode nonclassicality.  Nonclassicality can be
represented by a lack of proper probability distribution, in other
words, by the existence of singularities or negative values in the
Glauber-Sudarshan distribution~\cite{Mandel}. The density operator $\rho$ of a quantum state can be represented by the Glauber-Sudarshan distribution $P(\alpha)$ as
\begin{equation}
  \rho = \int P(\alpha)|\alpha\rangle\langle\alpha| d^2 \alpha,
\end{equation} 
where $|\alpha\rangle =
\exp(-|\alpha|^2/2)\sum_{n=0}^{\infty}\alpha^n/\sqrt{n!}|n\rangle$ is a
coherent state of amplitude $\alpha$. The decoherence in a vacuum implies the decrease of
amplitude with rate $\sqrt{\eta}$, i.e. $\alpha \rightarrow \sqrt{\eta}\alpha$ [see
the transformation in Eq.~(\ref{decoh2})] and it leads to rescaling of
the Glauber-Sudarshan distribution, $P(\alpha)\rightarrow
P(\alpha/\sqrt{\eta})/\eta$. Thus it is clear that the decoherence in a
vacuum does not completely destroy the nonclassicality of the
single-mode field. A similar conclusion can be arrived at for a multimode system.  However, for entanglement,
which is the nonclassicality in inter-mode correlations, such a
conclusion does not always hold. That is, the entanglement can vanish
even though nonclassicality still persists during
decoherence. 

As an example of the situation, when entanglement prevails for infinite time, consider an entangled state generated at the two outputs of a beam splitter by injecting  a nonclassical field and a vacuum into the respective input ports \cite{nonclassicality}. If both modes of the entangled state undergo the same amount of vacuum decoherence, the vacuum decoherence channels can be, formally, placed either before or after the beam splitter - the effects are equivalent. This implies that the generated entanglement is never completely lost in the vacuum decoherence, as the initial nonclassicality never is.  The situation does not change if the modes of the entangled state are affected by different amounts of vacuum decoherence. In this case, additional decoherence, an operation that can not increase entanglement, can be applied locally to transform the scenario to the previous one. 

However, in order to observe the dynamics of entanglement in general, we require reliable criteria of separability (or
measures of entanglement) to test the entanglement of a decohered state. A necessary and sufficient criterion derived in
Ref.~\cite{Horodecki1} states that a quantum state $\rho$ is separable if and
only if the image operator $(\openone\otimes\Lambda) \rho$ is a physical
state, for each positive map $\Lambda$ and an identity map $\openone$. Matrix
transposition is most often employed as a positive map $\Lambda$,
resulting in the well-known criterion of negative partial transposition (NPT)~\cite{NPT}. The NPT criterion is sufficient but not necessary for
arbitrary-dimensional systems to be entangled even though it is a
necessary and sufficient condition for small-dimensional systems, i.e., $2\times2$ and $2\times3$ systems. The
implementation of the NPT criterion also requires the complete knowledge of the full
density matrix and this can pose certain problems, especially for
infinite dimensional systems. Therefore, attempts have been made to use
the NPT as a base of construction for criteria that are easier to implement, such as the Simon criterion, employing second-order statistical moments, which is necessary
and sufficient for the class of Gaussian states~\cite{Simon}. However, to accommodate for a wider set of entangled states it was necessary to employ higher statistical moments of annihilation and creation operators, as in \cite{Agarwal, Shchukin, Hillery1}. 

To investigate whether and when the entanglement of a two-mode field
vanishes by vacuum decoherence, we use the Shchukin-Vogel criterion.
This criterion states that, for an entangled state, there exists a
hermitian matrix of moments, $M$, such that $\det M <0$ where $\det M$
is the determinant of $M$~\cite{Shchukin}. The matrix elements, $M_{ij}$
are certain moments of annihilation and creation operators for the two
modes $a$ and $b$,
\begin{eqnarray}\label{Melement}
  M_{ij} = \Tr[\mathcal{M}_{ij}(a,a^{\dag},b,b^{\dag})  \rho], 
\end{eqnarray} 
where
\begin{eqnarray}
   \mathcal{M}_{ij}(a,a^{\dag},b,b^{\dag}) = a^{\dag i_2}a^{i_1} a^{\dag j_1} a^{j_2} b^{\dag
     j_4}b^{j_3} b^{\dag i_3} b^{i_4},
  \nonumber 
\end{eqnarray} 
$\rho$ is the density operator for a two-mode field, and $i_k$, $j_l$
are proper components of multi-indices $i
\equiv (i_1,i_2,i_3,i_4)$ and $j \equiv
(j_1,j_2,j_3,j_4)$~\cite{Shchukin}.

Consider a subset of the
Shchukin-Vogel hierarchy such that matrix elements $M_{ij}^{N_a}$
are formed entirely by normally ordered moments (NOM) for mode $a$. This implies that $M^{N_a}$ is a submatrix of
$M$, formed by moments of operators
\begin{eqnarray}\label{momentsno}
  \mathcal{M}_{ij}^{N_a}(a,a^{\dag},b,b^{\dag}) = a^{\dag i_1} a^{j_1} b^{\dag
    j_3}b^{j_2} b^{\dag i_2} b^{i_3},
\end{eqnarray} 
where the multi-indices are now $i \equiv (i_1,i_2,i_3)$ and $j \equiv
(j_1,j_2,j_3)$. Note that the subset of $\mathcal{M}^{N_a}$
discriminates a certain set of states to be entangled, still forming a sufficient criterion of entanglement. 
This particular set will be called the $a$-mode NOM class of entangled states. One can perform a similar
procedure for both modes $a$ and $b$ instead of the single mode, and
obtains the $ab$-mode NOM class of entangled states.

The quantum state
$\rho_{ab}(t)$ of the two-mode field at the interaction time
$t$ can be calculated  for an initial state
$\rho_{ab}(0)$ as in Eq.~(\ref{decoh1}) if one of the two modes
interacts with a vacuum environment, say mode $a$. Then the $a$-mode NOM matrix elements $M^{N_a}_{ij}(t)$ for  $\rho_{ab}(t)$ are obtained as
\begin{eqnarray}
   \label{eq:nommout}
   M^{N_a}_{ij}(t) &=&
   \Tr \left[\rho_{ab}(t)
     \mathcal{M}_{ij}^{N_a}(a,a^{\dag},b,b^{\dag}) \right] 
  \nonumber \\ 
  &=&  \Tr\left[\rho_{ab}(0)\langle 0|_r U_{ar}^{\dag}
     \mathcal{M}_{ij}^{N_a}(a,a^{\dag},b,b^{\dag}) U_{ar}|0\rangle_r \right] \nonumber \\
  &=&  \Tr\left[\rho_{ab}(0)  \mathcal{M}_{ij}^{N_a}(\sqrt{\eta}a,\sqrt{\eta}a^{\dag},b,b^{\dag})\right]  
\end{eqnarray}
and  the Shchukin-Vogel determinant, $\det M^{N_a}(t)$, is in
the form of
\begin{equation}
\label{eq:svcsmnomc}
\det M^{N_a}(t) = \det(H) \det(M^{N_a}(0)) \det(H),
\end{equation} 
where $M^{N_a}(0)$ is the $a$-mode NOM matrix of the initial state and the matrix $H$ is diagonal with various powers of $\sqrt{\eta}$.  This relation between $\det M^{N_a}(t)$ and
$\det M^{N_a}(0)$ in Eq.~(\ref{eq:svcsmnomc}) means that {\it any entangled
state belonging to the $a$-mode (or $ab$-mode) NOM class remains 
entangled by the vacuum decoherence in mode $a$ (or modes $a$ and
$b$)}. Note that in case of the single-mode decoherence the condition
is less strict, as it requires only operators corresponding to the
decohering mode to be normally ordered.  The same conclusion is also
true for all states proven inseparable by the criteria,
recently proposed by Hillery and Zubairy~\cite{Hillery1,Hillery2}. This
is because all the moments used in their hierarchies are normally
ordered and the value of any particular criterion $c(0)$ transforms under the
vacuum decoherence as $c(t) = \eta^n c(0)$, where
$n$ is the order of the member of the hierarchy, and the sign does not
change.

This condition also ensures that under the vacuum decoherence, the
entanglement of Gaussian states is never lost. This follows directly
from the realization that Duan's necessary and sufficient criterion for
separability of Gaussian states~\cite{Duan} is one normally ordered
criterion of the Shchukin-Vogel hierarchy~\cite{Shchukin}. Other
states whose entanglement behave in this way include 
coherent superposition states $|\alpha,\alpha\rangle \pm
|-\alpha,-\alpha\rangle$, $|\alpha,-\alpha\rangle \pm
|-\alpha,\alpha\rangle$ (normalisation omitted), as can be checked by
first-order conditions in~\cite{Hillery1}.

\petr{
So far, when considering single mode decoherence, we have assumed the second mode to be left completely undisturbed. However, we have found that this ideal situation is not strictly necessary to keep entanglement, as long as the decoherence in the second mode, represented by the coupling $\eta_b$ (as opposed to the $\eta_a$ of the first mode), is reasonably small, i.e. $1-\eta_b \ll 1$. In this case, the a-mode NOM operator (\ref{momentsno}) transforms under the decoherence as:
\begin{eqnarray}
& &\langle 0,0|_{r_a,r_b} U_{\mathrm{tot}}^{\dag}\mathcal{M}_{ij}^{Na}(a,a^{\dag},b,b^{\dag})U_{\mathrm{tot}}|0,0\rangle_{r_a,r_b} \nonumber \\
&=& \langle 0|_{r_b} U_{br_b}^{\dag}\mathcal{M}_{ij}^{Na}(\sqrt{\eta_a} a,\sqrt{\eta_a}a^{\dag},b,b^{\dag}) U_{br_b}|0\rangle_{r_b} \nonumber \\
&=& \mathcal{M}_{ij}^{Na}(\sqrt{\eta_a} a,\sqrt{\eta_a}a^{\dag},\sqrt{\eta_b}b,\sqrt{\eta_b} b^{\dag}) \nonumber \\
& & + (1-\eta_b)\tilde{\mathcal{M}}_{ij}(\sqrt{\eta_a} a,\sqrt{\eta_a}a^{\dag},b,b^{\dag}),
\end{eqnarray} 
where $U_{\mathrm{tot}} = U_{ar_a}\otimes U_{br_b}$.
For a small coupling constant $\eta_b$ we can neglect the second term and the relevant determinant is again of the form (\ref{eq:svcsmnomc}), now with the diagonal matrix $H$ being composed of various powers of $\sqrt{\eta_a}$ and $\sqrt{\eta_b}$. Therefore the entanglement of the initial state will not be completely lost as long as the decoherence of the not-normally ordered mode is weak enough.
}

As a side note, let us ponder a while about the nature of the robustness against vacuum decoherence. We will argue that it has nothing to do with the unidirectionality of the energy flow, the fact that energy can be only lost in the process, but rather with the "coherence" of the evolution. To support this, let us imagine an environment where all the modes are in a coherent state $|\alpha\rangle$, the density matrix of the system after the decoherence being, in analogy with (\ref{decoh1}), 
\begin{equation}\label{coh1}
 \rho_{s}(t)= \Tr_r[U_{ar}
  \left( \rho_s(0)\otimes
    |\alpha\rangle_r {}_r\langle \alpha| \right) U_{ar}^{\dag}].
\end{equation}   
Writing the coherent state with help of the displacement operator \cite{Barnett}, $|\alpha\rangle = D(\alpha)|0\rangle$, 
$D(\alpha) = \exp( a^{\dag}\alpha - a\alpha^*)$, allows us to use the relation $U_{ar}D_r(\alpha)U_{ar}^{\dag} = D_a(-\alpha\sqrt{1-\eta})D_r(\alpha\sqrt{\eta})$ and to write the equation (\ref{coh1}) in the form
\begin{eqnarray}
\rho_{s}(t) = D_a(-\alpha e^{i\phi}\sqrt{1-\eta})\times \nonumber \\ 
\Tr_r [U_{ar} \rho_{s}(0)\otimes|0\rangle_{rr}\langle 0|U_{ar}^{\dag}]D_a(-\alpha e^{i\phi}\sqrt{1-\eta})^{\dag}.
\end{eqnarray}  
In terms of entanglement, this is clearly the same result as in the case of vacuum environment, as the only difference is in a single mode displacement, which is irrelevant to entanglement of the state. Therefore, any states that are robust against vacuum decoherence are also robust when  the decoherence is caused by a passive interaction with coherent environment.

\begin{figure}
\centerline{\psfig{figure=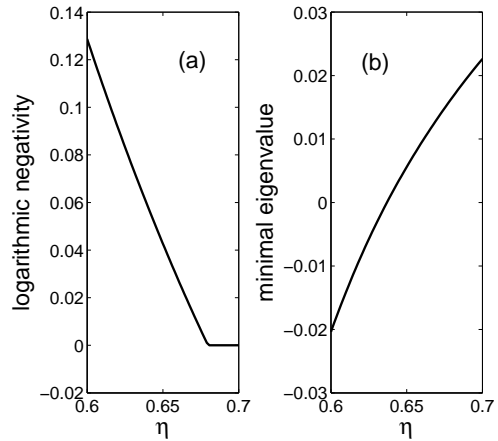,width=0.85\linewidth}}
\caption{The logarithmic negativity, $(a)$ and the minimal eigenvalue, $(b)$ of the state obtained by inverse decoherence with parameter $\eta$.}\label{LN}
\end{figure}

\begin{figure}
\centerline{\psfig{figure=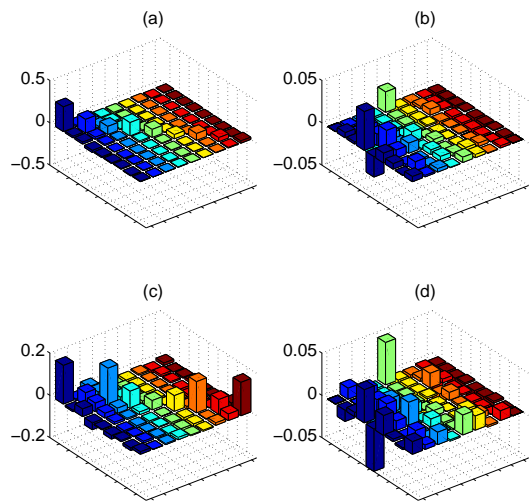,width=0.95\linewidth}}
\caption{ (Color online) The real and imaginary parts of the particular separable density matrix under consideration are depicted as $(a)$ and $(b)$, respectively. The density matrix of the (still physical) state affected by the inverse decoherence map with $\eta = 0.64$ is shown in $(c)$ (real part) and $d$ (imaginary part). The logarithmic negativity of this state is $E_{\mathcal{N}} \approx 0.06$.   }\label{statefig}
\end{figure}

Finally, as far as the authors are concerned, there has been no report on finite time disentanglement caused by vacuum decoherence for systems other than qubits \cite{Yu,Santos}. Although this is implied by geometric considerations in \cite{Cuhna}, using the existence of both entangled and separable states in a neighbourhood of the vacuum state, it is not shown whether evolution trajectories leading to finite time disentanglement really exist. Alternatively, attempts at specific analysis have to deal with a lack of necessary criteria of entanglement for generic high dimensional Hilbert space (HS), dealing only with lower bounds of entanglement \cite{Lastra} and again only suggesting that finite time disentanglement may take place. Here  we present another way of showing a possibility of disentanglement. 
It relies on using the inverse of the decoherence map  (\ref{decoh3}),
\begin{equation}\label{decohinv} 
\rho_{s,\mathrm{in}} = \sum_{n=0}^{\infty} (-1)^n L_n \rho_{s,\mathrm{out}} L_n^{\dag},
\end{equation}    
with
\begin{equation}
L_n = \sqrt{\eta}^{-a^{\dag}a}\left(\frac{a \sqrt{1-\eta}e^{i\phi}}{\sqrt{\eta}}\right)^n \frac{1}{\sqrt{n!}}.
\end{equation} 
This map is not positive and, if applied to an arbitrary state, it may result in an unphysical state. If the resulting state is physical, though, it is exactly the state that will decohere into the initial one. It is now possible to consider an arbitrary separable state of the form \begin{equation}
\rho_{\mathrm{sep}} = \sum_{k} p_k \rho_a^{(k)} \otimes \rho_b^{(k)},
\end{equation} 
where $\sum_k p_k=1$, and apply the inverse decoherence map (\ref{decohinv}) on one or both the subsystems of the state. If the resulting state is physical, commonly used entanglement criteria such as NPT can be used to decide upon entanglement of the state. As an example we have considered a bipartite state spanning HS of the dimension 3. The state was subject to inverse decoherence (\ref{decohinv}), with $\phi=0$ and varying parameter $\eta$, \petr{affecting both subsystems of the state}. Looking at Fig.~\ref{LN}, it can be seen that although the state becomes nonphysical around $\eta = 0.64$, it is NPT entangled roughly when $\eta = 0.68$. Here, the entanglement is quantified by means of logarithmic negativity \cite{logneg}, $E_{\mathcal{N}} = \log_2\parallel \rho^{PT}\parallel$, where $\parallel . \parallel$ denotes the trace norm and $\rho^{PT}$ stands for partially transposed density matrix of the state. Thus it can be seen that the finite time disentanglement caused by the vacuum decoherence is not limited to qubit systems. The particular state under discussion can be found in Fig.~\ref{statefig}.     

\section{Qubit systems}
Let us consider the vacuum decoherence in a bipartite qubit system. Although
the two-dimensional HS of qubits can be seen just as a
subspace of general infinite-dimensional space of continuous variables
with base states limited to ``vacuum" and ``single photon", it is an
important area of physics on its own, describing, for example, spin
systems and two-level atoms in cavities. We are going to use base states
$|0\rangle$ and $|1\rangle$ to be consistent throughout the paper, but
the states could also be denoted as $|g\rangle$ and $|e\rangle$
(ground and excited states of an atom) or $|\uparrow\rangle$ and
$|\downarrow\rangle$ (spin up, spin down). The noise model
(\ref{decoh3}) corresponds to decoherence by amplitude damping which, in
case of atoms in cavity, is caused by spontaneous emission of photon by
the excited atom. Of course, all operators $K_n$ for $n\geq2$ transform
the density matrix into zero.
 
In the qubit case, the NPT criterion serves as a necessary and
sufficient condition for entanglement and therefore conclusive results
about the dynamics of entanglement under the decoherence can be obtained. 
This allows us to find specific conditions for disentanglement and compare them to those obtained with help of the formalism presented so far.

Although the non-positivity of partially transposed density matrix
serves as a necessary and sufficient criterion for separability, it is
useful to quantify the entanglement by means of Wootters' concurrence~\cite{concurrence}
\begin{equation}\label{conc}
  C(\rho) = \max[0, \sqrt{\lambda_1}-\sqrt{\lambda_2}-\sqrt{\lambda_3}-\sqrt{\lambda_4}],
\end{equation}  
where $\lambda_i$ are eigenvalues of the matrix $\sqrt{\rho}
(\sigma_y\otimes\sigma_y) \rho^* (\sigma_y\otimes\sigma_y)\sqrt{\rho}$
arranged in decreasing order. Here,  $\sigma_y$ stands for the off-diagonal
pure imaginary Pauli matrix. The value of concurrence ranges from $C=0$
for separable states to $C=1$ for maximally entangled qubit states
containing one e-bit of entanglement. In the following we shall omit the notion of maximum, using the term concurrence  for $C = \sqrt{\lambda_1}-\sqrt{\lambda_2}-\sqrt{\lambda_3}-\sqrt{\lambda_4}$, but bearing in mind that it has good meaning only when positive.

Let us suppose a pure state of the general form
\begin{equation}\label{pure}
  |\psi\rangle = \alpha |0 0\rangle+\beta|01\rangle +\gamma|10\rangle+\delta|11\rangle.
\end{equation} 
The concurrence of this state is $C = 2|\alpha\delta - \beta\gamma|$ and
the state is initially entangled if $C>0$. If we let both particles of this
system decohere according to (\ref{decoh3}) with the same decoherence coefficient $\eta$, the decohered density matrix
would be\\
\begin{equation}
  \rho_{out} = \sum_{\mu=1}^{4} K'_{\mu}|\psi\rangle\langle \psi|K_{\mu}'^{\dag},
\end{equation} 
where the Kraus operators can be expressed in matrix
form
\begin{eqnarray}
  K'_1 &=& \left( 
    \begin{array}{cc}
      1 & 0 \\ 
      0 & \sqrt{\eta}
    \end{array}
  \right) \otimes \left(
    \begin{array}{cc}
      1 & 0 \\ 
      0 & \sqrt{\eta}
    \end{array}
  \right), \nonumber \\
  K'_2 &=& \left( 
    \begin{array}{cc}
      1 & 0 \\ 
      0 & \sqrt{\eta}
    \end{array}
  \right) \otimes \left(
    \begin{array}{cc}
      0 & \sqrt{1-\eta} \\ 
      0 & 0
    \end{array}
  \right), \nonumber \\
  K'_3 &=& \left( 
    \begin{array}{cc}
      0 & \sqrt{1-\eta} \\ 
      0 & 0
    \end{array}
  \right) \otimes \left(
    \begin{array}{cc}
      1 & 0 \\ 
      0 & \sqrt{\eta}
    \end{array}
   \right), \nonumber \\
  K'_4 &=& \left( 
    \begin{array}{cc}
      0 & \sqrt{1-\eta} \\ 
      0 & 0
    \end{array}
  \right) \otimes \left(
    \begin{array}{cc}
      0 & \sqrt{1-\eta} \\ 
      0 & 0
    \end{array}
  \right).
\end{eqnarray}
\medskip Applying these, we can find the eigenvalues required for
(\ref{conc}) to be
\begin{eqnarray}
  \lambda_1 &=&
  \eta\left[|\alpha\delta-\beta\gamma|+\sqrt{|\alpha\delta-\beta\gamma|^2+|\delta|^4(1-\eta)^2}\right]^2,\nonumber \\ 
  \lambda_2 &=&
  \eta\left[|\alpha\delta-\beta\gamma|-\sqrt{|\alpha\delta-\beta\gamma|^2+|\delta|^4(1-\eta)^2}\right]^2,\nonumber \\ 
  \lambda_3 &=& \lambda_4 = \eta^2(1-\eta)^2|\delta|^4
\end{eqnarray} 
and the concurrence is readily obtained as
\begin{equation}
  C^{(2)} = 2\eta\left(|\alpha\delta-\beta\gamma|-(1-\eta)|\delta|^2\right),
\end{equation} 
where the superscript $(2)$ denotes that we are considering decoherence of
both particles of the system.  This clearly shows that for pure
states of the form (\ref{pure}) with
$|\alpha\delta-\beta\gamma|>|\delta|^2$ the entanglement is, under the
vacuum decoherence, never completely lost \cite{Carvalho}. 
There is an alternative way to arrive at this conclusion, using the result in Section~\ref{sec:iii}. It follows from realization that when detecting entanglement of any pure bipartite qubit state using the Shchukin-Vogel criteria, the value of determinant composed of moments (\ref{momentsno}), with indices $(1,0,0)$, $(0,0,1)$ and $(1,0,1)$, which are normally ordered in both modes, reads $D_1 = |\delta|^2(|\delta|^4 - |\alpha\delta - \gamma\delta|^2)$. Therefore, for states with $\delta \neq 0$, the condition obtained is the same as in the case of concurrence. If $\delta = 0$ it is possible to consider a matrix of moments $(0,0,0)$, $(1,0,0)$ and $(1,0,1)$ whose determinant is $D_2 = -|\beta|^4|\gamma|^2$. We should keep in mind that, although it is not the case for these examples, this approach requires the state to be defined in the first two dimensions of general HS. 

However, it may be also interesting to study what happens if the same initial state is affected by decoherence of only one of the subsystems. In this case it shows that the entanglement of \emph{all} pure qubit states is not completely lost in a finite time, suggesting an inherent difference between single- and two-mode decoherence. To show this, the needed Kraus matrices are
\begin{equation}
  K''_1 = \left(
    \begin{array}{cc}
      1 & 0 \\ 
      0 & \sqrt{\eta}
    \end{array}
  \right), \quad
  K''_2 = \left(
    \begin{array}{cc}
      0 & \sqrt{1-\eta} \\ 
      0 & 0
    \end{array}
  \right)
\end{equation} 
with relevant eigenvalues
\begin{equation}
  \lambda_1 = 4\eta|\alpha\delta-\beta\gamma|^2, \quad \lambda_2=\lambda_3=\lambda_4 = 0,
\end{equation}
giving the concurrence, $C^{(1)}= 2\sqrt{\eta}|\alpha\delta-\beta\gamma|$, which is always positive.  The
superscript $(1)$ denotes the single-particle decoherence.   Therefore, single-particle vacuum decoherence never causes initial pure state to disentangle completely. Again, the result in Section~\ref{sec:iii} can be used as an alternative: the determinant of a $4\times 4$ matrix composed of moments (\ref{momentsno}) with indices $(1,0,0)$, $(0,0,1)$, $(1,0,1)$ and $(1,1,1)$ is equal to $D_3 = -|\delta|^4 |\alpha\delta-\beta\gamma|^2$ and is negative for all pure states if  $\delta \neq 0$. If $\delta = 0$  a matrix of moments $(0,0,0)$, $(1,0,0)$ and $(1,0,1)$ with determinant $D_2 = -|\beta|^4|\gamma|^2$ can be used again.  
Note, this interesting result suggests that given the option between decoherence of both subsystems and stronger decoherence of
only one of them, the latter possibility can be more considerate of entanglement. However, it is only for initially pure states,  when this property of single-particle vacuum decoherence always manifests, see \cite{impure} for an example. 

\petr{ As in the previous section, this conclusion needs not to be limited to cases where only one of the subsystems is affected by decoherence. Consider a situation, where the two coupling constants $\eta_a$ and $\eta_b$ are different. In this case one may find the concurrence to be 
\begin{equation}\label{conc2}
C'^{(2)} = 2\sqrt{\eta_a \eta_b}(|\alpha\delta - \beta\gamma| - \sqrt{(1-\eta_a)(1-\eta_b)}|\delta|^2)
\end{equation} 
}\petr{
and from here set the condition for disentanglement as $(1-\eta_a)(1-\eta_b) > |\alpha\delta-\beta\gamma|^2/|\delta|^4$. If $1-\eta_b < |\alpha\delta-\beta\gamma|/|\delta|^4$, even if the subsystem $a$ is approaching full decoherence, i.e. $\eta_a \rightarrow 0$, entanglement can still survive. Moreover, for a fixed value of the product $\eta_a\eta_b$ we can clearly see that the degree of entanglement in Eq.(\ref{conc2}) is minimized when $\eta_a = \eta_b$. The time dependant couplings, $\eta_a$ and $\eta_b$, depend on the quality of the channel and the travelling time. We can conclude that, the more the couplings are unbalanced, the more likely is for the entanglement to survive.
}

\section{Conclusions}
We have analysed the evolution of bipartite entangled quantum states
under the decoherence caused by passive linear coupling with a vacuum
environment. For general, continuous-variable, quantum states we have
found a class of states, whose entanglement will, under the vacuum
decoherence, never completely disappear. We have also shown that this result holds also when decoherence by passive interaction with coherent environment is considered. Furthermore, by means of the inverse decoherence map we have demonstrated that finite time disentanglement is not limited to qubit systems.

For qubit systems, the existence of necessary and sufficient criteria of entanglement allowed us to explicitly find the conditions of entanglement preservation under both single- and two-mode decoherence and to compare those with the results obtained with help of the general approach laid out in Section~\ref{sec:iii}. We have found a good agreement between them, which shows that it is possible to gain some insight into the qubit systems when treating them as a part of the general HS. 

\petr{
Furthermore, for a pure two-qubit state, when only one subsystem is affected by the decoherence, the entanglement is never completely lost. When both subsystems are affected, a finite time disentanglement may occur, but even in this case a difference in decoherence couplings will hinder this process. 
}
There can also be a practical implication of this finding in quantum communication tasks aimed at distributing entanglement between distant parties. We can consider two elementary scenarios: distributing the both parts of the system via noisy channels or sending just one part of the system via noisy channel with double the noise. In light of our result it is apparent that the first scenario can lead to complete disentanglement while the second one always preserves at least some entanglement, leaving the possibility of entanglement purification opened. 
\petr{ 
Keep in mind that this approach is beneficial even when it is impossible to shield single subsystem from decoherence completely, as the entanglement is still more likely to survive if the decoherence couplings are strongly unbalanced.
}
Note, that this reasoning can also be applied to infinite dimensional HS states whose entanglement can be verified only by criteria based on moments normally ordered in just one of the subsystems. 

\medskip
\emph{Note added:} Since our submission, it has been shown that for a two-qubit state the entanglement reduction under a noisy channel acting on one of the subsystems is independent on the initial state but completely determined by the channel's action on the maximally entangled state \cite{Konrad}. This can be seen as a generalisation of our finding about entanglement of a pure qubit state which is never completely lost under the single channel decoherence.

\medskip

We acknowledge the financial support of the UK EPSRC. P.M. acknowledges
support from the European Social Fund.  JL appreciates the financial support provided by the Korean
Research Foundation Grant funded by the Korean
Government (MOEHRD) (KRF-2005-041-C00197).


\begin{thebibliography}{99}
\bibitem{Barnett} S. M. Barnett and P. M. Radmore, Methods in theoretical quantum optics (Oxford, 1997).

\bibitem{EPR} A. Einstein, B. Podolsky, and N. Rosen, Phys. Rev. {\bf
    47}, 777 (1935).

\bibitem{Bell64} J. S. Bell, Physics (Long Island City, N.Y.) {\bf 1},
  195 (1964).

\bibitem{cryptography} For overview see, for example, N. Gisin, G.
  Ribordy, W. Tittel, and H. Zbinden, Rev. Mod. Phys. {\bf 74}, 145
  (2002).

\bibitem{teleportation} C. H. Bennett, G. Brassard, C. Crepeau, R. Jozsa,
  A. Peres, and W. K. Wootters, Phys. Rev, Lett. {\bf 70}, 1895 (1993);
  L. Vaidman, Phys. Rev. A {\bf 49}, 1473 (1994); S. L. Braunstein and
  H. J. Kimble, Phys. Rev. Lett {\bf 80}, 869 (1998).

\bibitem{computing} For overview see, for example, T. Spiller, W. Munro,
  S. Barrett, P. Kok, Cont. Phys. {\bf 46}, 407 (2005).

\bibitem{nonclassicality} M. S. Kim, W. Son, V. Bu\v{z}ek, and P. L.
  Knight, Phys. Rev. A 65, 032323 (2002).  J. K. Asb\'{o}th, J.
  Calsamiglia, and Helmut Ritsch, Phys. Rev. Lett. {\bf 94}, 173602
  (2005).
  
\bibitem{nonclassicality1}J. S. Ivan, N. Mukunda and R. Simon, quant-ph/0603255 (2006).

\bibitem{nonclassicality2} C. T. Lee, Phys. Rev. A {\bf 44}, R2775
  (1991).

\bibitem{Duan} L.-M. Duan, G. Giedke, J. I. Cirac, and P. Zoller, Phys.
  Rev. Lett. {\bf 84}, 2722 (2000); J. Lee, M. S. Kim and H. Jeong, \pra {\bf 62}, 032305 (2000).

\bibitem{Yu} T. Yu, and J. H. Eberly, Phys. Rev. Lett. {\bf 93}, 140404
  (2004). T. Yu and J. H. Eberly,  Phys. Rev. Lett. {\bf 97}, 140403 (2006). 

\bibitem{Santos} M. F. Santos, P. Milman, L. Davidovich, and N.
  Zagury, Phys. Rev. A {\bf 73}, 040305(R) (2006).

\bibitem{Almeida} M. P. Almeida, F. de Melo, M. Hor-Meyll, A. Salles, S.
  P. Wallborn, P. H. Souto Ribiero, and L. Davidovich, arXiv:
  quant-ph/0701184 (2007).

\bibitem{lee04} J. Lee, I. Kim, D. Ahn, H. McAneney, and M. S. Kim, \pra
  {\bf 70}, 024301 (2004); H. McAneney, J. Lee, and M. S. Kim, {\em
    ibid.} {\bf 68}, 063814 (2003); M. S. Kim and N. Imoto, {\em ibid.}
  {\bf 52}, 2401 (1995); H. Fearn and R. Loudon, Opt. Commun. {\bf 64},
  485 (1987).

\bibitem{carmichael} H. Carmichael, An Open Systems Approach to Quantum Optics (Springer, 1993, Berlin)

\bibitem{Mandel} L. Mandel, Phys. Scrip. {\bf T12}, 34 (1986).

\bibitem{glauber} For overwiev see, for example, U. Leonhardt, {\em
    Measuring the quantum state of light}, Cambridge University Press
  (2005).

\bibitem{Horodecki1} M. Horodecki, P. Horodecki, and R. Horodecki, Phys.
  Lett A {\bf 223}, 1 (1996).

\bibitem{NPT} A. Peres, Phys. Rev. Lett {\bf 77}, 1413 (1996).

\bibitem{Simon} R. Simon, Phys. Rev. Lett. {\bf 84}, 2726 (2000).

\bibitem{Agarwal}
 G. S. Agarwal and A. Biswas, New J. Phys. {\bf 7}, 350 (2005). 

\bibitem{Shchukin} E. Shchukin, and W. Vogel, Phys. Rev. Lett. {\bf 95},
  230502 (2005).

\bibitem{Hillery1} M. Hillery, and M. S. Zubairy, Phys. Rev. Lett. {\bf
    96}, 050503 (2006).

\bibitem{Hillery2} M. Hillery, and M. S. Zubairy, Phys. Rev. A {\bf 74},
  032333 (2006).

\bibitem{logneg}
G. Vidal, R. F. Werner, Phys. Rev. A {\bf 65}, 032314 (2002); 
J. Lee, M. S. Kim, Y. J. Park, and S. Lee, J. mod. Opt \textbf{47}, 2151 (2000).

\bibitem{concurrence} W. K. Wootters, Phys. Rev. Lett. {\bf 80}, 2245
  (1998).

\bibitem{Carvalho}
A. R. R. Carvalho, M. Busse, O. Brodier, C. Viviescas, and A. Buchleitner, Phys. Rev. Lett. {\bf 98}, 190501 (2007).

\bibitem{Cuhna}
M. O Terra Cuhna, New J. Phys. {\bf 9}, 237 (2007).

\bibitem{Lastra}
F. Lastra, G. Romero, C. E. L\'opez, M. Fran\c{c}a Santos, and J. C. Retamal, Phys. Rev. A {\bf 75}, 062324 (2007).

\bibitem{impure}
Consider mixed states of form
\begin{equation}
\rho_A = \left(
    \begin{array}{cccc}
      v & 0 & 0 & 0 \\ 
      0 & w & z & 0 \\ 
      0 & z^* & x  &0  \\ 
      0 & 0  & 0 & y
    \end{array}
  \right), \nonumber
\end{equation} 
where the rows and columns correspond to base states $(|00\rangle$,
$|01\rangle$, $|10\rangle$, $|11\rangle$).  
The state is physical if $|z|^2\leq wx$ and  entangled if $|z|^2>vy$. After single-particle decoherence, the concurrence of the state becomes
\begin{equation}
C_{A}^{(1)} = 2\sqrt{\eta}\left(|z_A| - \sqrt{vy+xy(1-\eta)}\right), \nonumber
\end{equation} 
and the state will remain entangled indefinitely only if $|z|^2>(v+x)y$. This result can be also obtained with use of the general result in Section~\ref{sec:iii}.

\bibitem{Konrad}
T. Konrad, F. de Melo, M. Tiersch, C. Kasztelan, A. Arag\~ao, and A. Buchleitner, Nature Physics \textbf{4}, 99 (2008).

\end{thebibliography}
\end{document}